\documentclass[preprint,showpacs,preprintnumbers,amsmath,amssymb]{revtex4}

\def\undersim#1{\setbox9\hbox{${#1}$}{#1}\kern-\wd9\lower
    2.5pt \hbox{\lower\dp9\hbox to \wd9{\hss $_\sim$\hss}}}
\usepackage{amssymb}
\usepackage{amsmath}
\usepackage{graphicx}

\def\undersim#1{\setbox9\hbox{${#1}$}{#1}\kern-\wd9\lower
    2.5pt \hbox{\lower\dp9\hbox to \wd9{\hss $_\sim$\hss}}}

\def\mr{{\mathbf r}}

\def\mr{{\mathbf r}}

\def\mk{{\mathbf k}}

\begin{document}

\title{Predictions for squeezed back-to-back correlations of $\phi\phi$ and
$K^+K^-$ in high-energy heavy-ion collisions by event-by-event hydrodynamics}

\author{Yong Zhang$^1$}
\author{Wei-Ning Zhang$^{1,\,2,}$\footnote{wnzhang@dlut.edu.cn}}
\affiliation{$^1$School of Physics and Optoelectronic Technology,
Dalian University of Technology, Dalian, Liaoning 116024, China\\
$^2$Department of Physics, Harbin Institute of Technology, Harbin,
Heilongjiang 150006, China}


\begin{abstract}
We calculate the squeezed back-to-back correlation (BBC) functions of $\phi
\phi$ and $K^+K^-$ for heavy-ion collisions at RHIC and LHC energies, using ($2+1$)-dimensional hydrodynamics with fluctuating initial conditions.  The
BBC functions averaged over event-by-event calculations for many events for
the hydrodynamic sources are smoothed as a function of the particle momentum.
For heavy-ion collisions of Au+Au at $\sqrt{s_{NN}}=200$ GeV, the BBC functions
are larger than those for collisions of Pb+Pb at $\sqrt{s_{NN}}=2.76$ TeV.
The BBC of $\phi\phi$ may possibly be observed in peripheral collisions at the
RHIC and LHC energies.  It is large for the smaller sources of Cu+Cu collisions
at $\sqrt{s_{NN}}=200$ GeV.


\end{abstract}

\maketitle

\section{Introduction}
In the late 1990s, it was shown \cite{AsaCso96,AsaCsoGyu99} that the mass
modification of the particles in the hot and dense hadronic sources can
lead to a squeezed back-to-back correlation (BBC) of boson-antiboson pairs
in high-energy heavy-ion collisions.  This BBC caused by the interactions
in medium is different from the pure quantum statistical correlations
between the bosons with different isospins \cite{AndPluWei91}, which
are negligible in high-energy heavy-ion collisions because the Fourier
transformation of source-density, $|{\tilde \rho}({\vec 0},m)|$, is very
small even for the lightest meson \cite{Gyu79,AsaCsoGyu99}.
Since it is associated with the source medium, the investigations of the
BBC may possibly provide another way for people to understand the thermal
and dynamical properties of the hadronic sources formed in high-energy
heavy-ion collisions, in addition to particle yields and spectra.

The BBC function is defined as \cite{AsaCso96,AsaCsoGyu99}
\begin{equation}
\label{BBCf}
C(\mk,-\mk) = 1 + \frac{|G_s(\mk,-\mk)|^2}{G_c(\mk,\mk) G_c(-\mk,-\mk)},
\end{equation}
where $G_c(\mk_1,\mk_2)$ and $G_s(\mk_1,\mk_2)$ are the chaotic and squeezed
amplitudes, respectively,
\begin{equation}
G_c(\mk_1,\mk_2)=\sqrt{\omega_{\mk_1}\omega_{\mk_2}}\,\langle
a^\dagger_{\mk_1} a_{\mk_2} \rangle,
\end{equation}
\begin{equation}
G_s(\mk_1,\mk_2)=\sqrt{\omega_{\mk_1}\omega_{\mk_2} }\,\langle
a_{\mk_1} a_{\mk_2}\rangle,
\end{equation}
where $a_\mk$ and $a^\dagger_\mk$ are the annihilation and creation operators 
of the free boson with momentum $\mk$ and mass $m$, $\langle \cdots \rangle$ 
indicates the ensemble average.  For a homogeneous source with volume $V$ and 
temperature $T$, the BBC function can be written as \cite{AsaCsoGyu99}
\begin{equation}
\label{BBCf1}
C(\mk,-\mk) = 1 + \frac{V\,|c_{\mk}\,s_{\mk}^*\,n_{\mk} +c_{-\mk}\,s_{-\mk}^*
\,(n_{-\mk}+1)|^2}{V\,[n_1(\mk)\,n_1(-\mk)]},
\end{equation}
where $c_\mk$ and $s_\mk$ are the coefficients of Bogoliubov transformation 
between the creation (annihilation) operators of the quasiparticle in medium 
with modified mass $m_*$ and the free observed particle 
\cite{AsaCso96,AsaCsoGyu99}, $n_\mk$ is boson distribution,  
\begin{equation}
n_{\mk}=\frac{1}{\exp(\Omega_{\mk}/T)-1},~~~(\Omega_\mk=\sqrt{\mk^2+m_*^2}), 
\end{equation}
and 
\begin{equation}
n_1(\mk)=|c_{\mk}|^2\,n_{\mk}+|s_{-\mk}|^2(n_{-\mk}+1).
\end{equation}
The BBC function $C(\mk,-\mk)$ will be 1 if there is no mass modification.  
However, for a finite mass modification, $\delta m^2 =m^2-m^2_*$, $f_{\mk}
\sim \delta m^2/(4\mk^2)$ as $|\mk|\to\infty$, and the BBC function will 
increase with increasing particle momentum \cite{AsaCsoGyu99}, $C(\mk,-\mk)
\sim 1+1/|s_{-\mk}|^2\sim 1+\mk^4/(\delta m^2/4)^2$.

For hydrodynamic sources, with the formula derived by Makhlin and Sinyukov
\cite{MakhSiny}, the chaotic and squeezed amplitudes can be expressed as
\cite{AsaCsoGyu99,Padula06,Padula10,YZHANG15a}
\begin{eqnarray}
\label{Gchydro}
&& G_c({\mk_1},{\mk_2})\!=\!\int \frac{d^4\sigma_{\mu}(r)}{(2\pi)^3}
K^\mu_{1,2}\, e^{i\,q_{1,2}\cdot r}\,\! \Bigl\{|c'_{\mk'_1,\mk'_2}|^2\,
n'_{\mk'_1,\mk'_2}~~~~~~\nonumber \\
&& \hspace*{19mm}
+\,|s'_{-\mk'_1,-\mk'_2}|^2\,[\,n'_{-\mk'_1,-\mk'_2}+1]\Bigr\},
\end{eqnarray}
\begin{eqnarray}
\label{Gshydro}
&& G_s({\mk_1},{\mk_2})\!=\!\int \frac{d^4\sigma_{\mu}(r)}{(2\pi)^3}
K^\mu_{1,2}\, e^{2 i\,K_{1,2}\cdot r}\!\Bigl\{s'^*_{-\mk'_1,\mk'_2}
c'_{\mk'_2,-\mk'_1}~~~~~\nonumber \\
&& \hspace*{18mm}
\times n'_{-\mk'_1,\mk'_2}+c'_{\mk'_1,-\mk'_2} s'^*_{-\mk'_2,\mk'_1}
[n'_{\mk'_1,-\mk'_2} + 1] \Bigr\}.
\end{eqnarray}
Here $d^4\sigma_{\mu}(r)$ is the four-dimension element of freeze-out
hypersurface, $q^{\mu}_{1,2}=k^{\mu}_1-k^{\mu}_2$, $K^{\mu}_{1,2}=
(k^{\mu}_1+k^{\mu}_2)/2$, and $\mk_i'$ is the local-frame momentum
corresponding to $\mk_i~(i=1,2)$.  The local quantities, $c'_{\mk'_1, 
\mk'_2}$, $s'_{\mk'_1,\mk'_2}$, and $n'_{\mk'_1,\mk'_2}$ are the 
coefficients of Bogoliubov transformation and boson distribution 
associated with the particle pair (see in Ref. \cite{YZHANG15a}).  

In Eq. (\ref{Gshydro}), the factor $e^{2iK_{1,2}\cdot r}$ is equal to
$e^{2i\omega_{\mk}t}$ for $\mk_1=\mk$, $\mk_2=-\mk_1=-\mk$.  So the BBC
function $C(\mk,-\mk)$ is sensitive to the temporal distribution of
the freeze-out points \cite{AsaCsoGyu99,Padula06,Padula10,YZHANG15a,Kno11}.
Recent research \cite{YZHANG15a} indicates that the BBC functions for
the hydrodynamic sources with Gaussian initial-energy distributions
exhibit oscillations as a function of the particle momentum because
of the sharp falls of temporal freeze-out distributions at long times.
Investigating the BBC behavior and predicting its effect in high-energy
heavy-ion collisions based on more realistic models is of great interest
\cite{YZHANG15a}.  In this work, we investigate the BBC functions for the
hydrodynamic sources with event-by-event fluctuating initial conditions
(FIC).  We use ($2+1$)-dimensional hydrodynamics with the HIJING
\cite{WanGyu91} FIC and the equation of state s95p-PCE \cite{She10} to
describe the source evolution as in Ref. \cite{HuZha15}, and investigate
the BBC functions of $\phi\phi$ and $K^+K^-$ for heavy-ion collisions of
Au+Au at $\sqrt{s_{NN}}=$ 200 GeV at the Relativistic Heavy Ion Collider
(RHIC) and Pb+Pb at $\sqrt{s_{NN}}=$ 2.76 TeV at the Large Hadron
Collider (LHC), respectively.  Our investigation indicates that the BBC
functions averaged over event-by-event calculations for many events for
the hydrodynamic sources with the FIC are smoothed and without the
oscillations appearing in the BBC functions for the hydrodynamic sources
with Gaussian initial-energy distributions \cite{YZHANG15a}.
For heavy-ion collisions at the RHIC and LHC energies, the BBC of $\phi
\phi$ may possibly be observed in peripheral collisions. The investigations
for the BBC functions of $\phi\phi$ for Cu+Cu collisions at $\sqrt{s_{NN}}
=200$ GeV indicate that the BBC is large for the smaller sources.

The rest of this paper is organized as follows.  In Sec. II, we give a
brief review on the relativistic hydrodynamic model used in this work,
and discuss the space-time distributions of particle freeze-out points
for the hydrodynamic sources with the FIC.  In Sec. III, we present the
BBC functions of $\phi\phi$ and $K^+K^-$ for heavy-ion collisions of
Au+Au and Pb+Pb at the RHIC and LHC energies, respectively.
The dependences of the BBC functions on the azimuthal angle and
pseudorapidity of the particles are also investigated in this section.
In Sec. IV, we investigate the source space-time distributions and the
BBC functions of $\phi$ meson for the smaller sources in Cu+Cu collisions
at $\sqrt{s_{NN}}=200$ GeV.  Finally, a summary and conclusions of this
paper are given in Sec. V.

\section{($2+1$)-dimensional hydrodynamic sources}
Relativistic hydrodynamics has been extensively applied in high-energy
heavy-ion collisions.  In this work, we use the ideal relativistic
hydrodynamics in $(2+1)$ dimensions to describe the transverse expansion
of the particle-emitting sources formed in ultrarelativistic heavy-ion
collisions, and adopt the Bjorken boost-invariant hypothesis \cite{Bjo83}
for the longitudinal evolution of the sources.
The hydrodynamic equations of motion for the sources with zero net
baryon density are from the local conservation of energy-momentum
\cite{Ris98,KolHei03}.  Under the assumption of Bjorken longitudinal
boost invariance, we need only to solve the transverse equations of
motion in the $z=0$ plane \cite{HuZha15}, and the hydrodynamic solutions
at $z\ne0~(v^z=z/t)$ can be obtained by the longitudinal boost invariance
\cite{Bay83,Gyu97}.

Assuming the local equilibrium of system is reached at time $\tau_{0}$,
we construct the initial energy density of the hydrodynamic source at
$z=0$, by using the AMPT code \cite{Lin05} in which the HIJING is used
for generating the initial conditions, as \cite{Gyu97,Pan12}
\begin{eqnarray}
\label{E-ic}
\epsilon(\tau_0,x,y;z=0)=K\sum_{\alpha}\frac{p_{\perp\alpha}}{\tau_0}
\frac{1}{2\pi\sigma_0^2}\,\exp\left\{-\frac{[x-x_{\alpha}(\tau_{0})]^{2}
+[y-y_{\alpha}(\tau_{0}
)]^2}{2\pi\sigma_0^2}\right\}.
\end{eqnarray}
Here $p_{\perp\alpha}$ is the transverse momentum of parton $\alpha$
in the fluid element at $(x,y)$, $x_{\alpha}(\tau_{0})$ and $y_{\alpha}
(\tau_{0})$ are the transverse coordinates of the parton at $\tau_{0}$,
$\sigma_0$ is a transverse width parameter, and $K$ is a scale factor
which can be adjusted to fit the experimental data of produced hadrons
\cite{Pan12}.  The initial velocity of the fluid element is then
determined by the initial energy density and the average transverse
momentum of the partons in the element.

As in Ref \cite{HuZha15}, we solve the hydrodynamic equations
numerically by the HLLE scheme and the Sod's operation splitting
method \cite{HLLE,Ris98,Ris9596,ZhaEfa,Zha04,Yu08Yin12,Sod77}.
In the calculations, we use the equation of state s95p-PCE \cite{She10},
which combines the hadron resonance gas at low temperatures and the
lattice QCD results at high temperatures, and take the parameters
$\sigma_0=0.6$ fm, $\tau_0=$ 0.6 and 0.4 fm/$c$ for the heavy-ion
collisions at the RHIC and LHC, respectively.  With these parameter
values the hydrodynamic results of transverse momentum spectra and
elliptic flow of identical pion and charged hadrons are consistent
with the experimental dada at the RHIC and LHC \cite{Pan12,HuZha15}.

For hydrodynamic sources with a Bjorken cylinder, the four-dimension
element of freeze-out hypersurface can be written as
\begin{equation}
d^4\sigma_{\mu}(r)=f_{\mu}(\tau, \mr_{\perp}, \eta)\,d\tau d^2
\mr_{\perp} d\eta,
\end{equation}
where $\tau$, $\mr_{\perp}$, and $\eta$ are the proper time, transverse
coordinate, and space-time rapidity of the element.  The function $f_{\mu}
(\tau,\mr_{\perp},\eta)$ is related to the freeze-out mechanism that is
considered, and $K^{\mu}_{1,2}f_{\mu}(\tau,\mr_{\perp},\eta)$ corresponds
to the source distributions of proper time and space in the calculations
[see Eqs. (\ref{Gchydro}) and (\ref{Gshydro})].  In this work we assume
that $\phi$ and $K$ mesons are frozen out at fixed temperatures $T_f=$
140 and 160 MeV, respectively, and use the AZHYDRO technique
\cite{Kol00,KolRap03,KolHei03} to calculate the freeze-out hypersurface
element.

\begin{figure}[htbp]
\includegraphics[scale=0.46]{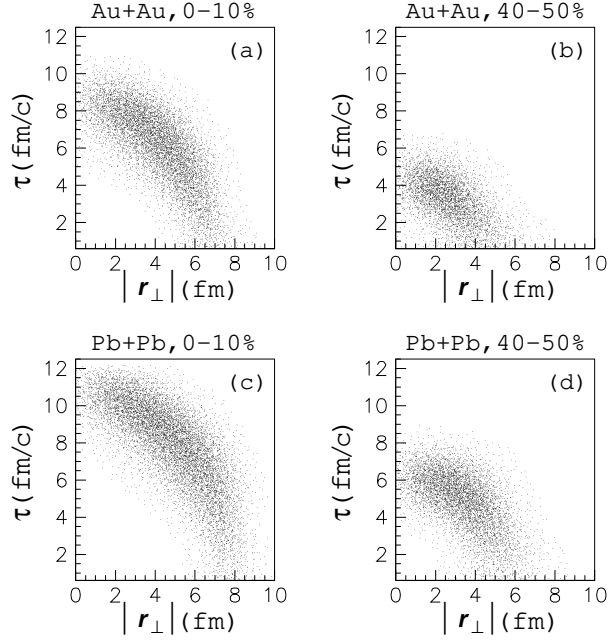}
\vspace*{-5mm}
\caption{Distributions of the freeze-out points of $\phi$ meson in
the $z=0$ plane for central and peripheral collisions of Au+Au
at $\sqrt{s_{NN}}=200$ GeV and Pb+Pb at $\sqrt{s_{NN}}=2.76$ TeV,
plotted with 2000 events. }
\label{disfop}
\end{figure}

\begin{figure}[htbp]
\includegraphics[scale=0.46]{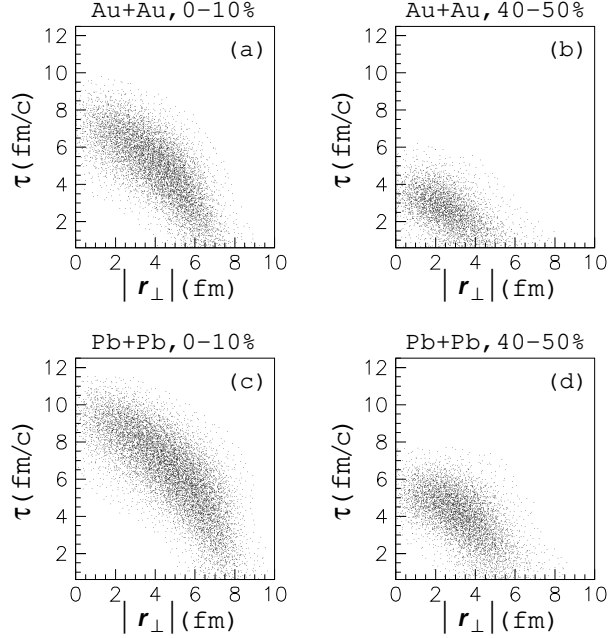}
\vspace*{-5mm}
\caption{Distributions of the freeze-out points of $K$ meson in
the $z=0$ plane for the central and peripheral collisions as in Fig.
\ref{disfop}, plotted with 2000 events. }
\label{disfok}
\end{figure}

We plot in Fig. \ref{disfop} the space-time distributions of the freeze-out
points, $k^{\mu}f_{\mu}(\tau,\mr_{\perp},\eta)$, of $\phi$ meson in the $z=0$
plane for central and peripheral heavy-ion collisions of Au+Au at $\sqrt{
s_{NN}}=200$ GeV at the RHIC and Pb+Pb at $\sqrt{s_{NN}}=2.76$ TeV at the LHC.
Here, the event number for both the RHIC and LHC collisions is 2000.
The regions of impact parameter $b$ for the denoted centralities 0--10\% and
40--50\% are taken to be 0--4.2 and 9.0--10.2 fm \cite{STAR05}, respectively.
One can see that the width of the distributions increases with decreasing
collision centrality and increasing collision energy.  For the hydrodynamic
sources with FIC, the freeze-out distributions are more dispersive compared
to the hydrodynamic sources with the Gaussian initial conditions
\cite{YZHANG15a}.  In Fig. \ref{disfok}, we plot the space-time distributions
of the freeze-out points of $K$ meson in the $z=0$ plane for the central and
peripheral collisions as in Fig. \ref{disfop}.  The widths of the freeze-out
distributions of $K$ meson are smaller than those of $\phi$ meson because of
the higher freeze-out temperature of $K$ meson.

\begin{figure}[htbp]
\includegraphics[scale=0.60]{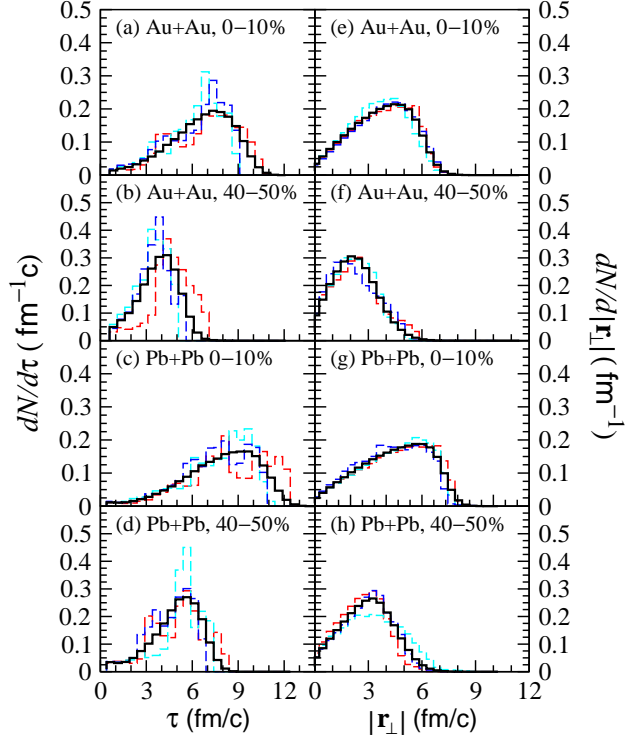}
\vspace*{-2mm}
\caption{(Color online) Normalized distributions of time and transverse
coordinate of the $\phi$ freeze-out points in the $z=0$ plane for the
heavy-ion collisions as in Fig. \ref{disfop}.  The thin dashed lines
are for different single events and the thick solid lines are for 2000
events. }
\label{disfop_tr}
\end{figure}

In Fig. \ref{disfop_tr}, we show the normalized distributions of time
and transverse coordinate of $\phi$ freeze-out points in the $z=0$ plane
for different single events (thin dashed lines) and 2000 events (thick
solid lines), for heavy-ion collisions of Au+Au and Pb+Pb at the RHIC
and LHC energies, respectively.  One can see that the distributions for
single events are fluctuated and these fluctuations are smoothed in the
distributions for 2000 events.

\section{BBC functions for ${\rm\bf Au+Au}$ at RHIC and ${\rm\bf Pb+Pb}$
at LHC}
Because of the fluctuations of the space-time coordinates of freeze-out
points, the BBC functions for single events are fluctuated.  However,
these fluctuations can be smoothed in the BBC functions averaged over
event-by-event calculations for many events as,
\begin{equation}
C(\mk,-\mk)=\frac{\frac{1}{N_E}\sum_{i=1}^{N_E}\left[G_{ci}(\mk,\mk)G_{ci}(-\mk,-\mk)
+|G_{si}(\mk,-\mk)|^2\right]}{\frac{1}{N_E}\sum_{i=1}^{N_E}G_{ci}(\mk,\mk)G_{ci}(-\mk,
-\mk)},
\end{equation}
where $N_E$ is the total event number.

\begin{figure}[htbp]
\includegraphics[scale=0.65]{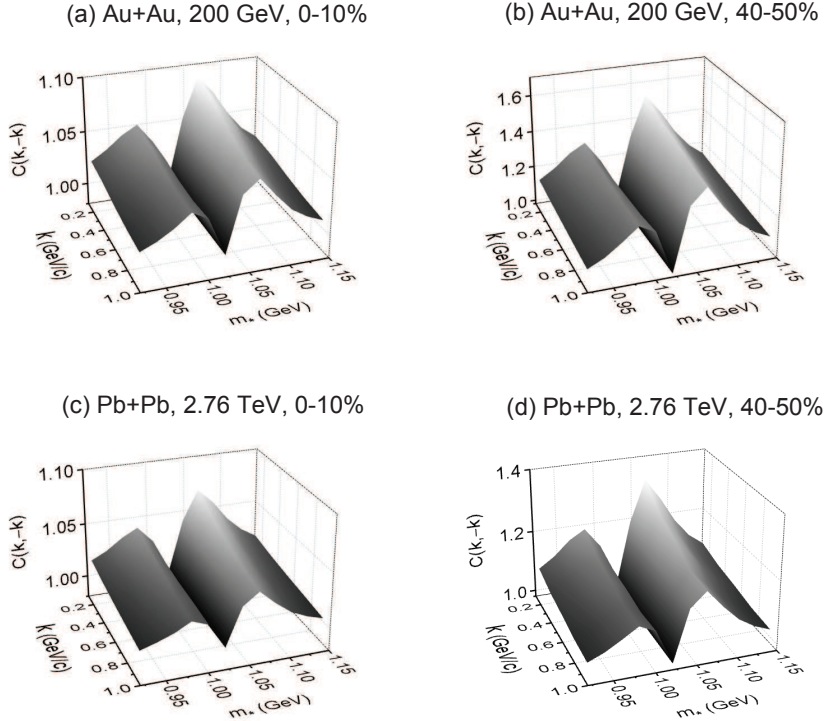}
\vspace*{-5mm}
\caption{BBC functions of $\phi\phi$ averaged over event-by-event
calculations for 2000 events for central and peripheral collisions
of Au+Au at $\sqrt{s_{NN}}=200$ GeV and Pb+Pb at $\sqrt{s_{NN}}=
2.76$ TeV. }
\label{BBCFP_RL}
\end{figure}

We show in Fig. \ref{BBCFP_RL} the BBC functions of $\phi\phi$ averaged
over event-by-event calculations for central and peripheral collisions
of Au+Au at $\sqrt{s_{NN}}=200$ GeV and Pb+Pb at $\sqrt{s_{NN}} =2.76$
TeV.  In the calculations of the BBC functions in this work, the total
event number $N_E$ for each kind of heavy-ion collision is 2000, and the
space-time rapidity of source is taken to be $|\eta|\leq 1$.  Unlike the
oscillations appearing in the BBC functions for the hydrodynamic sources
with Gaussian distributions of initial-energy density \cite{YZHANG15a},
the BBC functions for the hydrodynamic sources with FIC vary smoothly with
respect to the particle momentum.  This is because the oscillations in the
single-event BBC functions are canceled each other in the BBC functions
averaged over event-by-event calculations for many events.
From Fig. \ref{BBCFP_RL} one can see that the BBC functions for
peripheral collisions are larger than those for central collisions.
Also, it can be seen that the BBC functions for collisions at the
RHIC energy are larger than those for collisions at the LHC energy.

In Eqs. (\ref{Gchydro}) and (\ref{Gshydro}), the factors $e^{iq_{1,2}\cdot 
r}$ and $e^{2iK_{1,2}\cdot r}$ are equal to 1 and $e^{2i\omega_{\mk}t}$,
respectively, for $\mk_1=\mk$, $\mk_2=-\mk_1=-\mk$.  The BBC function 
$C(\mk,-\mk)$ defined with Eq. (\ref{BBCf}) is proportional to the square 
of the Fourier transformation of the source temporal distribution with the 
weight $[s'^*_{-\mk'_1,\mk'_2} c'_{\mk'_2,-\mk'_1} n'_{-\mk'_1,\mk'_2} + 
c'_{\mk'_1,-\mk'_2} s'^*_{-\mk'_2,\mk'_1}(n'_{\mk'_1,-\mk'_2}+1)]$, and the 
effects of the source spatial distribution in the numerator and denominator 
in Eq. (\ref{BBCf}) may cancel out partially.  So, the values of the BBC 
function are mainly related to the source temporal distribution, particle 
mass, mass shift, and source temperature.  Because we use the same 
freeze-out temperature for $\phi$ meson in the calculations for the 
collisions with different centralities and energies, the main reason 
of the dependences of the $\phi$ BBC function on the collision centrality 
and energy (see Fig. \ref{BBCFP_RL} for fixed $k$ and $m_*$) is that the 
temporal distribution of the source is narrower in peripheral collisions 
and becomes wider with increasing collision energy (see Fig. \ref{disfop}).

\begin{figure}[htbp]
\vspace*{5mm}
\includegraphics[scale=0.65]{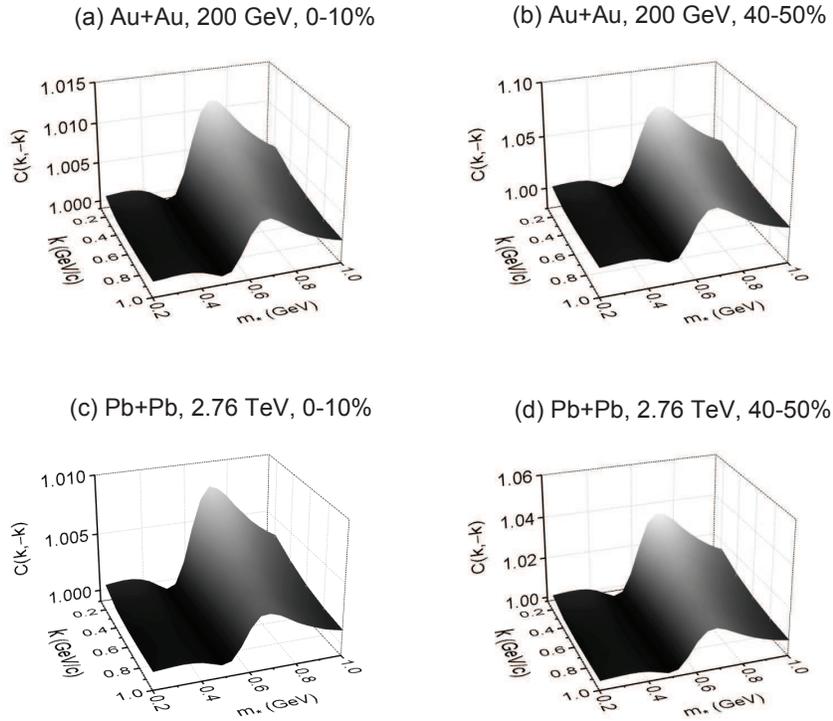}
\vspace*{-5mm}
\caption{BBC functions of $K^+K^-$ averaged over event-by-event
calculations for 2000 events for central and peripheral collisions
of Au+Au at $\sqrt{s_{NN}}=200$ GeV and Pb+Pb at $\sqrt{s_{NN}}=
2.76$ TeV. }
\label{BBCFK_RL}
\end{figure}

We show in Fig. \ref{BBCFK_RL} the BBC functions of $K^+K^-$ averaged
over event-by-event calculations for central and peripheral collisions
of Au+Au at $\sqrt{s_{NN}}=200$ GeV and Pb+Pb at $\sqrt{s_{NN}}=2.76$ TeV.
As compared to the results of the BBC functions of $\phi\phi$, the BBC
functions of $K^+K^-$ are much smaller.  This is mainly because the higher
freeze-out temperature and smaller mass of $K$ meson lead to larger values
of $e^{-k^{\mu}u_{\mu}/T_f}$ in the calculations of BBC functions.

\begin{figure}[!htbp]
\vspace*{5mm}
\includegraphics[scale=0.6]{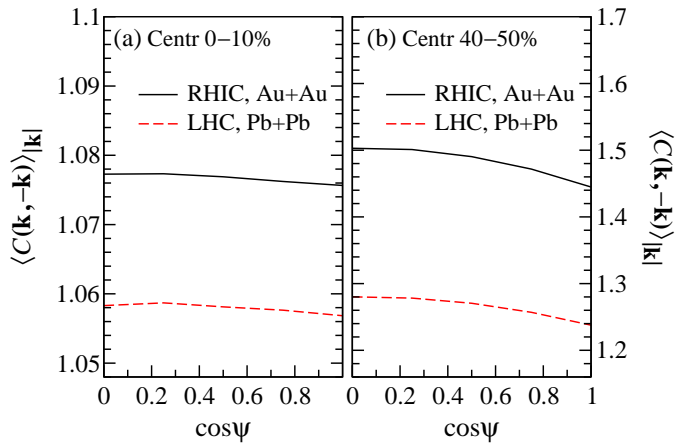}
\vspace*{-2mm}
\caption{(Color online) Dependence of the average BBC functions of
$\phi\phi$ on the cosine of the particle azimuthal angle $\psi$ for
the collisions at the RHIC and LHC energies.  Here, $m_*$ is taken
as 1.05 GeV and the momentum region averaged is 0--1 ${\rm GeV}\!/\!c$. }
\label{BBCFP_RL_b}
\end{figure}

For anisotropic sources, the anisotropic source velocity may lead
to the dependence of the BBC function on the direction of particle
momentum \cite{YZHANG15,YZHANG15a}.  We show in Fig. \ref{BBCFP_RL_b}
the dependence of the average BBC functions of $\phi\phi$, $\langle
C(\mk,-\mk)\rangle_{|\mk|}$, on the cosine of the particle azimuthal
angle $\psi$ for the collisions at the RHIC and LHC energies.  Here,
$m_{\!*}$ is taken as 1.05 GeV, corresponding approximately to the
middle between the valley and peak of the BBC function (see Fig.
\ref{BBCFP_RL}), and the momentum region averaged is 0--1 ${\rm GeV}
\!/\!c$.  The BBC functions decrease with increasing $\cos\psi$ more
rapidly for peripheral collisions because the source transverse
velocities are more anisotropic in this case \cite{YZHANG15,YZHANG15a}.

\begin{figure}[!htbp]
\vspace*{5mm}
\includegraphics[scale=0.6]{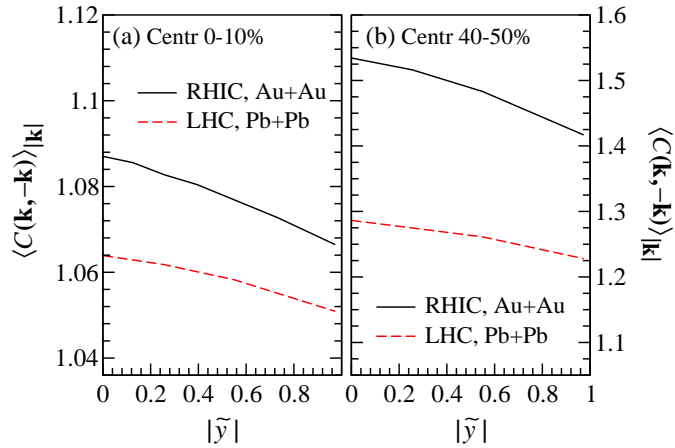}
\vspace*{-2mm}
\caption{(Color online) Dependence of the average BBC functions of
$\phi\phi$ on the particle pseudorapidity for the collisions at the
RHIC and LHC energies.  Here, $m_*$ is taken as 1.05 GeV and the
momentum region averaged is 0--1 ${\rm GeV}\!/\!c$. }
\label{BBCFP_RL_y}
\end{figure}

We show in Fig. \ref{BBCFP_RL_y} the dependence of the average BBC
functions of $\phi\phi$, $\langle C(\mk,-\mk)\rangle_{|\mk|}$, on
the particle pseudorapidity for the collisions at the RHIC and LHC
energies.  Here, $m_*$ is taken as 1.05 GeV and the momentum region
averaged is 0--1 ${\rm GeV}\!/\!c$.  One can see that the BBC functions
decrease with increasing $|\widetilde y|$.
This is because the average source longitudinal velocity is higher than
the average source transverse velocity for hydrodynamic sources with
Bjorken longitudinal boost invariance, and the higher longitudinal
velocity leads to larger average values of $e^{-k^{\mu}u_{\mu}/T_f}$
at $|\widetilde y|=1$ than at $|\widetilde y|=0$ \cite{YZHANG15,YZHANG15a}.

\section{BBC for ${\rm\bf Cu+Cu}$ collisions at ${\rm\bf \sqrt{s_{NN}}=
200\,GeV}$}
As we know squeezed BBC function is sensitive to the temporal distribution
of the particle freeze-out points.  The narrow temporal distribution
corresponding to a small hydrodynamic source may lead to a large BBC.
We are therefore motivated to investigate next the BBC for heavy-ion
collisions of Cu+Cu at $\sqrt{s_{NN}}=200$ GeV, which have smaller sources
compared to those in Au+Au and Pb+Pb collisions.

\begin{figure}[htbp]
\includegraphics[scale=0.50]{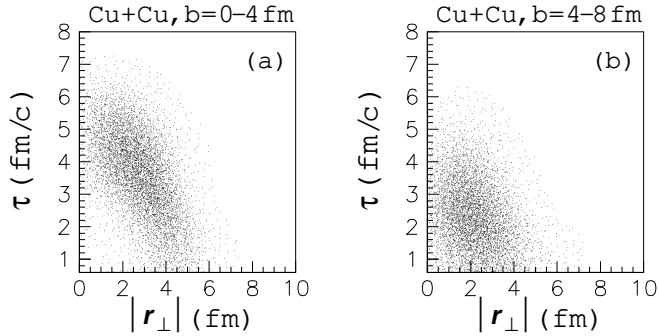}
\vspace*{-46mm}
\caption{Distributions of the freeze-out points of $\phi$ meson in the
$z=0$ plane for Cu+Cu collisions at $\sqrt{s_{NN}}=200$ GeV and with
the impact parameter $b=$ 0--4 and 4--8 fm.  The number of event is 2000. }
\label{disfop_cu}
\end{figure}

\begin{figure}[!htbp]
\vspace*{5mm}
\includegraphics[scale=0.75]{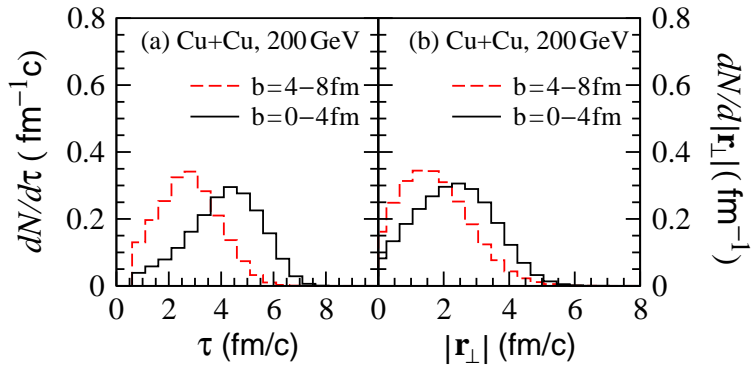}
\vspace*{-3mm}
\caption{(Color online) Normalized distributions of time and transverse
coordinate of the $\phi$ freeze-out points in the $z=0$ plane for
Cu+Cu collisions at $\sqrt{s_{NN}}=200$ GeV.  The solid lines are for
$b=$ 0--4 fm, and the dashed lines are for $b=$ 4--8 fm.  The number of
event is 2000. }
\label{disfop_tr_cu}
\end{figure}

In Figs. \ref{disfop_cu}(a) and \ref{disfop_cu}(b), we plot the space-time
distributions of the freeze-out points of $\phi$ meson in the $z=0$ plane
for Cu+Cu collisions at $\sqrt{s_{NN}}=200$ GeV with impact parameter
$b=$ 0--4 and 4--8 fm, respectively.  The event number for both the two
impact parameter regions is 2000.  One can see that the distributions
for Cu+Cu collisions are smaller compared to the distributions in Figs.
\ref{disfop}(a) and \ref{disfop}(b) for the Au+Au collisions.
In Figs. \ref{disfop_tr_cu}(a) and \ref{disfop_tr_cu}(b), we show the
normalized distributions of time and transverse coordinate of $\phi$
freeze-out points in the z = 0 plane for the Cu+Cu collisions.
The normalized time and transverse coordinate distributions are obtained
by projecting the space-time distributions in Fig. \ref{disfop_cu} to the
time and transverse coordinate axes, respectively.  One can see that the
widths of the distributions for peripheral collisions are obviously
narrower than those for central collisions.

\begin{figure}[htbp]
\vspace*{2mm}
\includegraphics[scale=0.70]{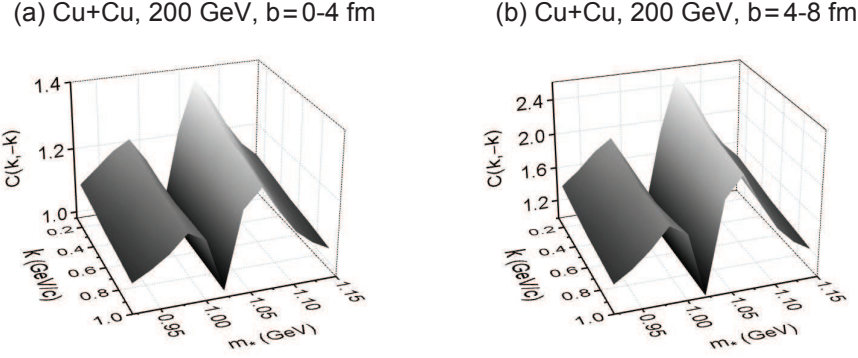}
\vspace*{-4mm}
\caption{BBC functions of $\phi\phi$ averaged over event-by-event
calculations for 2000 events for Cu+Cu collisions at $\sqrt{s_{NN}}
=200$ GeV and with $b=$ 0--4 and 4--8 fm. }
\label{BBCFP_cu}
\end{figure}

We plot in Figs. \ref{BBCFP_cu}(a) and \ref{BBCFP_cu}(b) the BBC functions
of $\phi\phi$ averaged over event-by-event calculations for 2000 events for
Cu+Cu collisions with impact parameter $b=$ 0--4 and 4--8 fm, respectively.
The BBC function for peripheral collisions is larger than that for central
collisions.  The peaks of the BBC functions reach about 1.3 and 2.5 for
central and peripheral collisions, respectively.  They are much higher
than the corresponding results for Au+Au collisions at the RHIC energy
and Pb+Pb collisions at the LHC energy, as shown in Fig. \ref{BBCFP_RL}.
The reason for this is mainly that the temporal distributions for the
smaller sources of Cu+Cu collisions are very narrow.

\begin{figure}[!htbp]
\vspace*{1mm}
\includegraphics[scale=0.60]{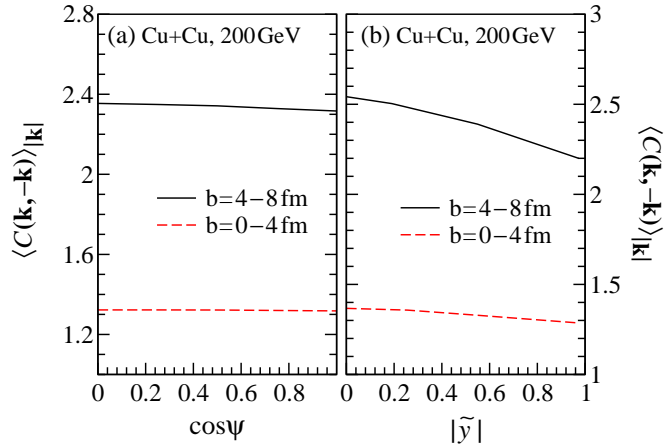}
\vspace*{-2mm}
\caption{(Color online) Dependences of the average BBC functions of
$\phi\phi$ on the cosine of the particle azimuthal angle and the
particle pseudorapidity for Cu+Cu collisions at $\sqrt{s_{NN}}=$
200 GeV.  Here, $m_*$ is taken as 1.05 GeV and the momentum region
averaged is 0--1 ${\rm GeV}\!/\!c$. }
\label{BBCFPhz_cu}
\end{figure}

In Figs. \ref{BBCFPhz_cu}(a) and \ref{BBCFPhz_cu}(b), we show the dependences
of the averaged BBC functions $\phi\phi$, $\langle C(\mk,-\mk)\rangle_{|\mk|}$,
on the cosine of the particle azimuthal angle $\psi$ and the particle
pseudorapidity for Cu+Cu collisions, respectively.  Here, $m_*$ is taken
as 1.05 GeV and the momentum region averaged is 0--1 GeV/$c$.
One can see that the average BBC functions almost independent of $\cos\psi$
for central collisions, and decrease slightly with increasing $\cos\psi$
for peripheral collisions because of the anisotropic transverse expansion
\cite{YZHANG15,YZHANG15a}.  As discussed in section III, the average BBC
functions for Cu+Cu collisions also decrease with the increasing
$|\widetilde y|$ like that for the Au+Au and Pb+Pb collisions.

\section{Summary and conclusions}
In the hot and dense hadronic sources formed in high-energy heavy-ion
collisions, the particle interactions in medium might lead to a squeezed
BBC of boson-antiboson pairs.  In this paper, we investigate the BBC
functions of $\phi\phi$ and $K^+K^-$ for heavy-ion collisions of Au+Au
at $\sqrt{s_{NN}}=200$ GeV at the RHIC and Pb+Pb at $\sqrt{s_{NN}}=2.76$
TeV at the LHC, using ($2+1$)-dimensional hydrodynamics with the FIC
generated by HIJING.  The investigations indicate that the BBC functions
averaged over event-by-event calculations for many events for the
hydrodynamic sources with the FIC are smoothed as a function of the
particle momentum.  This is different from the BBC functions for the
hydrodynamic sources with Gaussian initial-energy distributions, which
exhibit oscillations with respect to the particle momentum \cite{YZHANG15a}.
The BBC functions for collisions of Au+Au at the RHIC energy are larger
than those for collisions of Pb+Pb at the LHC energy, because the width
of source temporal distribution increasing with increasing collision
energy.  The average BBC function, $\langle C(\mk,-\mk)\rangle_{|\mk|}$,
exhibits a decrease with increasing the cosine of the particle azimuthal
angle for peripheral collisions because of the anisotropic source
transverse velocity.  Based on the calculations, we predict that the BBC
of $\phi\phi$ may possibly be observed in peripheral collisions of Au+Au
at the RHIC energy and Pb+Pb at the LHC energy.  The BBC of $\phi\phi$
is large for the smaller sources of Cu+Cu collisions at $\sqrt{s_{NN}}
=200$ GeV.  It is of great interest to investigate experimentally the 
BBC of $\phi\phi$ in high-energy heavy-ion collisions.

\begin{acknowledgments}
This research was supported by the National Natural Science Foundation
of China under Grant No. 11275037.
\end{acknowledgments}

\end{document}